\documentstyle[12pt]{article}
\newcommand{\newsection}{
\setcounter{equation}{0}
\section}

\def\appendix#1{
  \addtocounter{section}{1}
  \setcounter{equation}{0}
  \renewcommand{\thesection}{\Alph{section}}
 \section*{Appendix \thesection\protect\indent \parbox[t]{11.715cm} {#1}}
  \addcontentsline{toc}{section}{Appendix \thesection\ \ \ #1}
  }

\newcommand{\tr}[1]{\:{\rm tr}\,#1}
\newcommand{\Tr}[1]{\:{\rm Tr}\,#1}
\newcommand{\Sp}[1]{\:{\rm Sp}\,#1}
\def\e{\,{\rm e}\,}
\newcommand{\rf}[1]{(\ref{#1})}

\newcommand{\non}{\nonumber \\*}
\def\be{\begin{equation}}
\def\ee{\end{equation}}
\def\bea{\begin{eqnarray}}
\def\eea{\end{eqnarray}}
\def\const{{\rm const}}
\def\ll{\Lambda}
\def\d{\partial}
\def\D{\delta}
\def\q{\psi}
\def\bq{\bar{\psi}}
\def\dq{\psi^{\dagger}}
\def\g{\gamma}
\def\h{\eta}
\def\ep{\varepsilon}
\def\bh{\bar{\eta}}
\def\dh{\eta^{\dagger}}
\def\ud{u^{\dagger}}

\begin{document}
\begin{titlepage}
\begin{flushright}
hep-ph/9804276\\
April, 1998
\end{flushright}
\vspace{1.5cm}

\begin{center}
{\LARGE Renormalization of Schr\"odinger Equation} 
\\[.5cm]
{\LARGE and Wave Functional for Rapidly Oscillating Fields}
\\[.5cm]
{\LARGE in QCD}\\
\vspace{1.9cm}
{\large K.~Zarembo}\\
\vspace{24pt}
{\it Department of Physics and Astronomy,}
\\{\it University of British Columbia,}
\\ {\it 6224 Agricultural Road, Vancouver, B.C. Canada V6T 1Z1} 
\\ \vskip .2 cm
and\\ \vskip .2cm
{\it Institute of Theoretical and Experimental Physics,}
\\ {\it B. Cheremushkinskaya 25, 117259 Moscow, Russia} \\ \vskip .5 cm
E-mail: {\tt zarembo@theory.physics.ubc.ca/@itep.ru}
\end{center}
\vskip 2 cm
\begin{abstract}
Background field method is used to perform 
renormalization group transformations for Schr\"odinger 
equation in QCD. The dependence of the ground state wave functional 
on rapidly oscillating fields is found.   
\end{abstract}

\end{titlepage}

\setcounter{page}{2}

\newsection{Introduction}  

The usual applications of background field method include perturbative
calculations of Wilsonian effective action and of OPE coefficients, and
are commonly associated with path integral techniques. However, 
background field method can also be implemented in the Schr\"odinger 
representation \cite{lus83}. The Hamiltonian counterpart of the 
background field method was used, for example, in the soliton 
quantization \cite{raj89} and in a variational approach to the 
Yang-Mills vacuum \cite{hmvi98}.

The Schr\"odinger picture is seldom used
in field theory, but it may provide some complementary information
in comparison to the path integral approach, since only
in the Schr\"odinger 
representation the wave functions explicitly appear. The main 
motivation for this paper is to study the dependence of QCD vacuum wave
function on large-momentum modes of the fields. In the pure 
gauge theory this dependence was found \cite{zar98} by performing 
renormalization group transformations for the Schr\"odinger equation. 
We generalize this 
procedure to include quarks. Schr\"odinger representation for
fermions does not cause principal difficulties \cite{fj88}, moreover,
functional Schr\"odinger equation for fermions interacting with an
external gauge field, which arises in the background field method, 
was considered previously \cite{kw94}.  

Large-momentum modes in QCD are weakly coupled due to asymptotic freedom,
which makes it possible to find the solution of the Schr\"odinger
equation in a part of the configuration space corresponding to 
rapidly oscillating fields. The averaging of the Hamiltonian with the 
wave functional for large-momentum modes can be regarded a 
renormalization group transformation. The averaging over
rapidly oscillating fields was also used to study a lifting of classical
vacuum degeneracy by quantum corrections in some field theories 
\cite{lus83,ave}. However, other methods to exclude high-energy
modes from the Schr\"odinger equation exist \cite{oth}.
We use the background field technique 
because it allows to deal explicitly with the wave functional.

\newsection{Schr\"odinger equation for large-momentum modes}

The QCD Hamiltonian in the temporal gauge, $A_0=0$, is  
\begin{equation}\label{ham}
H=\int d^3x\,\left[\frac{g^2}{2}\,E_i^AE_i^A
+\frac{1}{4g^2}\,F_{ij}^AF_{ij}^A
-\sum_{f=1}^{N_f}\left(i\bq_f\g^iD_i\q_f-m_f\bq_f\q_f\right)
\right],
\end{equation}
where $F_{ij}^A=\partial _iA_j^A-\partial _jA_i^A+f^{ABC}A_i^BA_j^C$.
Gauge potentials $A_i^A(x)$ and electric fields $E_i^A(x)$
obey canonical commutation relations:
\begin{equation}\label{ccr}
[A_i^A(x),E_j^B(y)]=i\delta ^{AB}\delta _{ij}\delta (x-y).
\end{equation}
The quark fields satisfy anticommutation relations:
\be
\{\q^a_{\alpha f}(x),\q^{b\,\dagger}_{\beta f'}(y)\}
=\D^{ab}\D_{\alpha \beta}\D_{ff'}\D(x-y),
\ee
where $a$, $f$ and $\alpha$ are color, flavor and spinor indices,
respectively. We shall also use matrix notations for gauge
potentials: $A_i=A_i^AT^A$, where $T^A$ are anti-Hermitian generators
of $SU(N_c)$ normalized by $\tr T^AT^B=-\D^{AB}/2$, and the  
covariant derivative in the fundamental representation is $D_i=
\d_i+A_i$.

The wave functions of physical states are subject to Gauss'
law constraint:
\begin{equation}\label{gauss}
\left(D_iE_i^A+i\sum_f\dq_f T^A\q_f\right) \Psi =0.
\end{equation}
The covariant derivative acting on the electric fields is defined by
 $D_i^{AB}=\delta ^{AB}\partial _i+f^{ACB}A_i^C$.
The operator on the left hand side generates
gauge transformations, so the Gauss' law is the condition of the 
gauge invariance of physical states.

The renormalizability  guarantees that all UV divergencies can be 
removed from the Schr\"odinger equation in QCD  by standard 
counterterms \cite{sim81}. To perform renormalization group
transformations explicitly, we separate fast modes introducing a 
scale $\mu$, which is large enough for the running 
coupling $g^2(\mu)$ to be sufficiently small, and splitting the fields 
 in high- and low-energy components:
\bea\label{decom}
&&A_i\rightarrow A_i+ga_i,~~~~~E_i\rightarrow E_i+\frac{1}{g}\,e_i,
\non
&&\q_f\rightarrow\q_f+\eta_f,~~~~~\dq_f\rightarrow\dq_f+\dh_f.
\eea
We preserve the same notation for the low-energy components of the
fields which contain modes with momenta $p<\mu$ as for 
the full field variables.
The high-energy fields $a_i$, $e_i$, $\h_f$ and $\dh_f$ contain
only modes with momenta $\ll>p>\mu$. An upper bound on momenta is 
necessary for UV regularization. The large-momentum components of
gluon fields are rescaled for later convenience.

The low-energy fields are assumed to satisfy
classical equations of motion:
\be\label{cl}
D_jF_{ij}^A-ig^2\sum_f\bq_f\g^iT^A\q_f=0.
\ee
This assumption is conventional in the background field formalism. In 
fact, it is too restrictive, and a
more weak  condition that the
left hand side of eq.~\rf{cl} does not contain any large-momentum modes
will be sufficient in what follows. 

It is convenient to separate large-momentum 
components of fermion fields using the eigenmodes of the Dirac operator
in the background low-energy gluon field:
\be\label{dir}
h\equiv -i\g^0\g^iD_i+\g^0m,~~~~~h\,v_\ep=\ep v_\ep.
\ee
The low-energy quark fields $\q$ are then defined in such a way 
that they have 
zero projections on the eigenfunctions $v_\ep$ unless $|\ep|<\mu$. 
On the other 
hand, rapidly oscillating components of quark operators, $\h$, 
contain only 
high-energy harmonics $v_\ep$ with $\ll>|\ep|>\mu$.

After the decomposition \rf{decom} is substituted in the 
Hamiltonian, terms linear in fast variables vanish, and the Hamiltonian
splits in the parts describing low- and high-energy degrees of freedom. 
To the leading order in $g$, the
high-energy Hamiltonian is
quadratic in rapidly oscillating fields:
\be\label{ymham}
H_h=\int d^3x\,\left[\frac{1}{2}\,e_i^Ae_i^A+\frac{1}{2}\,
a_i^A(-D^2\D_{ij}-2F_{ij}+D_iD_j)^{AB}a_j^B
-\sum_{f}\left(i\bh_f\g^iD_i\h_f-m_f\bh_f\h_f\right)\right].
\ee
The Schr\"odinger representation is straightforward for the gluon part 
of this Hamiltonian --
the wave function is a functional of gauge potentials, and electric
fields act as functional derivatives: $e_i^A=-i\D/\D a_i^A$. 
The situation is more involved for quarks, because
canonical conjugation of a fermion field coincides with the Hermitian 
one. Hence, $\h$ and $\dh$ cannot be simultaneously considered 
as anticommuting c-numbers. It is possible to avoid the problem with 
Hermitian conjugation representing the quark operators in
the form \cite{fj88}:
\be\label{psiu}
\h_f(x)=\frac{1}{\sqrt{2}}\left(u_f(x)+\frac{\D}{\D\ud_f(x)}\right),
~~~~~\dh_f(x)=\frac{1}{\sqrt{2}}\left(\ud_f(x)
+\frac{\D}{\D u_f(x)}\right).
\ee
The wave functional in this representation 
depends on $u$ and $\ud$, which anticommute and are
complex conjugate to one another. The price for this  is
nonstandard inner product of the fermion wave functionals and
spurious enlargement of the Hilbert space in which quark operators act 
\cite{fj88}.

The Schr\"odinger equation for the Hamiltonian \rf{ymham} is
\be\label{sch}
\int d^3x\,\left[-\frac12\,\frac{\D^2}{\D a_i^2}+\frac12\,a_iL_{ij}a_j
+\frac12\sum_f\left(\ud_f+\frac{\D}{\D u_f}\right)h_f
\left(u_f+\frac{\D}{\D\ud_f}\right)\right]\Psi_h
=E_h\Psi_h,
\ee
where
\be\label{qf}
L_{ij}=-D^2\D_{ij}-2F_{ij}+D_iD_j
\ee
and $h$ is given by eq.~\rf{dir}. 
Low-energy variables in the Schr\"odinger equation \rf{sch} are
considered as external background fields. Since the Hamiltonian is
quadratic, the vacuum can be defined with the help of creation and
annihilation operators. The gluon vacuum is an empty state, while
the quark one is obtained by filling all negative-energy levels
in the Dirac sea. The gluon vacuum is described by the Gaussian
wave functional. The Schr\"odinger representation for fermions
interacting with an external gauge field was considered in 
Ref.~\cite{kw94}. Using the results of this paper, we obtain for
 the lowest-energy solution of Eq.~\rf{sch}: 
\be\label{ymwf}
\Psi_h=\exp\left[-\frac{1}{2}\,aL^{1/2}a+\sum_f\ud_f(P_f^--P_f^+)u_f
\right],
\ee
where the integration, as well as summation over color and over 
spatial indices is implied in the exponent.
The operators $P^-$ and $P^+$ are projectors on the spaces of negative
and of positive eigenmodes of the one-particle Hamiltonian \rf{dir}, 
respectively:
\be
P^-(x,y)=\sum_{\ep<-\mu}v^{\dagger}_\ep(x)v_\ep(y),~~~~~ 
P^+(x,y)=\sum_{\ep>\mu}v^{\dagger}_\ep(x)v_\ep(y).
\ee 

The wave functional \rf{ymwf} satisfies Gauss' law up to the two first
orders in $g$:
\be
\left(\frac{1}{g}\,D_ie_i^A+D_iE_i^A+f^{ABC}A_i^Be_i^C+iJ_0^A
\right)\Psi_h=0.
\ee
The first term on the left hand side generates transformations
\be\label{g1}
a_i\rightarrow a_i+D_i\omega.
\ee
The wave functional is invariant under these transformations because
$L_{ij}D_j\omega=0$ \cite{hmvi98}. Terms of
order $g^0$ in the Gauss' law generate gauge transformations of the
background fields, the large-momentum gluon variables transforming
as matter fields in the adjoint representation:
\be\label{g2}
A_i\rightarrow\Omega^{\dagger}(D_i+A_i)\Omega,
~~~~a_i\rightarrow\Omega^\dagger a_i\Omega.
\ee
The bosonic part of the wave functional is evidently gauge invariant.

The  Gauss' law for fermions requires 
more careful consideration. 
It appears that, when the representation 
\rf{psiu} is used for quark operators, not all gauge invariant states 
are annihilated by the Gauss' law operator with the straightforward 
definition of the quark current in eq.~\rf{gauss} 
\cite{fj88,kw94}, although this operator generates gauge 
transformation of $\psi$ and $\dq$. The paradox is related to the
enlargement of the fermion Hilbert space in $u$-representation 
\cite{fj88}. A proper definition  of the  
fermion current $J_0^A$ in $u$-representation is given in 
Ref.~\cite{kw94}. The operator defined in \cite{kw94} annihilates 
gauge invariant functionals 
of $u$ and $\ud$ and generates gauge transformations of fermion fields. 
Of course, this operator contains both high- and low-energy components
of the quark fields. But,
since we separate high-energy modes in a gauge-invariant 
way, the quark part of the wave functional is invariant under gauge 
transformations. 
Really, full (containing both high- and low-energy modes) quark 
operators and projectors transform as $\q(x)\rightarrow
\Omega^{\dagger}(x)\q(x)$, $P^\pm(x,y)\rightarrow\Omega^{\dagger}(x)
P^\pm(x,y)\Omega(y)$. Consequently, the high-energy  quark 
fields, being the projections of the full quark operators:
$\h=(P^-+P^+)\q$, also transform homogeneously: $\h(x)\rightarrow
\Omega^{\dagger}(x)\h(x)$, which ensures gauge
invariance of the wave functional. 

\newsection{Renormalization of Hamiltonian}

The fast modes can be excluded from consideration by writing the wave
functional in the form $\Psi=\Psi_h\Psi_l$, 
where $\Psi_l$ depends only on
low-energy fields. But the wave equation for $\Psi_l$ should be 
properly modified to account for zero-point energy of excluded rapidly 
oscillating degrees of freedom \cite{wil65}. 
The ground state energy of the Hamiltonian \rf{ymham},
\be\label{eff}
E_h=\frac{1}{2}\,\Tr L^{1/2}+\sum_f\Tr h_fP_f^-,
\ee
depends on the background fields and, 
thus, induces an effective
potential for low-energy variables:
$$
H_{\rm eff}=H+E_h,
$$
where $H$ is the bare Hamiltonian for slow degrees of freedom
which coincides in a form with \rf{ham}.
We assume that $\mu$
is sufficiently large for power-like
corrections in $1/\mu$ to be neglected, so that only UV 
divergent terms in the effective potential are
essential. The leading contribution is 
field-independent. It renormalizes zero-point energy 
and is not considered below. Since the remaining UV
divergencies can only renormalize the gauge coupling, 
the form of the
effective potential is clear from the outset, but it is instructive
to calculate it explicitly in order 
to check the consistency of our approach.

The first term in \rf{eff} 
can be calculated with the use of DeWitt-Seeley expansion for the 
operator $L$ given in Ref.~\cite{glno93}. The result is \cite{zar98}
\be\label{gl}
\frac12\,\Tr L^{1/2}=
\const-\frac{11N_c}{48\pi^2}\,\ln\frac{\ll^2}{\mu^2}\,\frac{1}{4}\,
\int d^3x\,F_{ij}^AF_{ij}^A+O(1/\mu^2).
\ee
Here we compute the second term. 

Assuming that $\mu\gg m_f$ for all 
$f$ we can neglect the mass term in the Dirac operator. Then the 
one-particle Hamiltonian \rf{dir} has
a symmetric spectrum, since it anticommutes with $\g^0$, and for any
eigenfunction $v_\ep$ with the eigenvalue $\ep$, 
$\g^0v_\ep$ is also an eigenfunction with the eigenvalue 
having an opposite sign, $-\ep$.
Thus,
\be\label{33}
\Tr hP^-=-\Tr hP^+=-\frac12\,\Tr(P^-+P^+)(h^2)^{1/2}.
\ee
It is more convenient to use a smooth proper-time cutoff 
instead of the sharp projection on the states with eigenvalues 
$\mu<|\ep|<\ll$. It means that in the heat kernel representation
for $(h^2)^{1/2}$,
\be\label{heatkforh}
(h^2)^{1/2}=\const-\frac{1}{2\pi^{1/2}}\,\int
\frac{d\tau}{\tau^{3/2}}\,\e^{-\tau h^2},
\ee
the integration over $\tau$ should range 
from $1/\ll^2$ to $1/\mu^2$. 
The heat kernel of 
the operator
\be
h^2=-D^2-\frac12\,\sigma^{ij}F_{ij},~~~~~\sigma^{ij}=
-\frac12\,[\g^i,\g^j]
\ee
at small $\tau$ can be expanded in local operators \cite{sch89}:
\be\label{dws}
\Sp\langle x|\e^{-\tau h^2}|x\rangle=\frac{4}{(4\pi\tau)^{3/2}}
\left(1+\frac{1}{12}\,\tau^2F_{ij}^AF_{ij}^A+O(\tau^3)\right),
\ee
where $\Sp$ denote the trace with respect to the spinor and to the 
color indices. Using eqs.~\rf{33}, \rf{heatkforh} and 
DeWitt-Seeley expansion \rf{dws}, we get:
\bea
\sum_f\Tr h_fP_f^-&=&\const+\frac{N_f}{2}\,\frac{1}{2\pi^{1/2}}\,  
\int\frac{d\tau}{\tau^{3/2}}\,\Tr\e^{-\tau h^2}
\non
&=&\const+\frac{N_f}{24\pi^2}\,
\ln\frac{\ll^2}{\mu^2}\,\frac{1}{4}\, 
\int d^3x\,F_{ij}^AF_{ij}^A+O(1/\mu^2).
\eea
This expression, together with the gluon contribution \rf{gl},
leads to the coupling constant renormalization:
\be
\frac{1}{g^2_{\rm eff}}=\frac{1}{g^2}-\frac{1}{8\pi^2}\left(
\frac{11N_c}{3}-\frac{2N_f}{3}\right)
\ln\frac{\ll}{\mu}
\ee
with the correct $\beta$-function.

\newsection{Discussion}

In principle, background field method allows to study not only
wave functional of the vacuum, but also that of excited states.
Low-energy excitations do not alter the wave functional $\Psi_h$
for rapidly oscillating fields, while high-energy excitations
correspond to some of the 
levels described by one-particle Hamiltonians \rf{dir} and \rf{qf}
being filled.

Background field method is perturbative in its nature and thus is
limited to the large-momentum modes of the 
fields. The complete ground state
wave functional in QCD including low-energy modes 
must be very sophisticated, but, perhaps,
it can be reasonably approximated by some variational 
ansatz\footnote{The variational approach to the 
ground state in the gauge theory recently was discussed in 
Refs.~\cite{kk95,hmvi98}}. Such ansatz ought to preserve basic
symmetries of QCD -- gauge, Lorenz and renormalization 
invariances. The equation \rf{ymwf} shows how should
the trial wave functional depend on large-momentum modes of the
fields in order to be compatible with the asymptotic freedom.
 
\subsection*{Acknowledgments}

This work was supported by NATO Science Fellowship and, in part, by
 CRDF grant 96-RP1-253,
 INTAS grant 96-0524,
 RFFI grant 97-02-17927
 and grant 96-15-96455 of the support of scientific schools.

\end{document}